\documentclass[12pt,a4paper]{article}

\usepackage{graphicx}

\usepackage{amssymb,amsmath,amsfonts}%,
\usepackage{axodraw,epsfig}

\def\,{\ifmmode\mskip\thinmuskip\else\leavevmode\thinspace\fi}

\def\be{\begin{equation}}
\def\ee{\end{equation}}
\def\bc{\begin{center}}
\def\ec{\end{center}}

\newcommand{\bea}{\begin{eqnarray}}
\newcommand{\eea}{\end{eqnarray}}

\begin{document}

%\hfill JINR, E2-2006-80, Dubna, 2006

\title{Enhancement of the $K \to \pi$ Transition by
the Instantaneous  Weak Interactions}
\medskip
\author{A.Z. Dubni\v{c}kov\'{a}$^{1},$ S. Dubni\v{c}ka$^{2},$
 V.N. Pervushin$^{3}$, M. Se\v cansk\'y$^{3,2}$ \\
\\
$^{1}${\small Department of Theoretical Physics, Comenius
University, Slovak
Republic}\\
$^{2}${\small  Institute of Physics, Slovak Academy of Sciences,
Bratislava,
Slovak Republic}\\
$^{3}${\small Joint Institute for Nuclear Research, 141980 Dubna,
Russia}}

\maketitle
\begin{center}
\end{center}

\vspace*{1cm}
\begin{abstract}
The hadrodynamics of  the  instantaneous weak interactions (IWI)
in Standard Model  is considered supposing the resonance nature of
the chiral hadronization of quark currents in QCD.
  Exploiting this supposition
 and QCD symmetries the IWI
 mechanism of enhancement of the $K \to
 \pi$ transition probability is obtained and
  relations between
 electroweak meson form factors and the amplitudes of the
 processes $K^+ \to \pi^+e^{+}e^{-}(\mu^{+}\mu^{-},\nu\bar{\nu})$
 and $K^+ \pi^-\to e^{+}e^{-}(\mu^{+}\mu^{-},\nu\bar{\nu})$ are established.
 We estimate possible parameters of the $K^+ \to \pi^+e^{+}e^{-}$
 decay rate in NA48/2 CERN experiment using the chiral
 perturbation theory.
\end{abstract}

%%\begin{titlepage}

\medskip

%\date{\empty}
 \maketitle

\vspace{-17cm}

\hfill Dubna, JINR, E-2006-80

\vspace{17cm}

%\tableofcontents

%\newpage

\section{Introduction}

The observation of kaon weak decays has been crucial for the
modern theory of particle physics \cite{mikulec}.
 It was accepted to  describe  weak decays  in the framework of
electroweak theory at the quark QCD level
 including  current vector boson weak interactions \cite{am,va}.
 However, a consistent theory at large distances of QCD is not yet
constructed up to now. Therefore, now  the most effective method
of analysis  of kaon decay physics
\cite{bvp,ecker,belkov01,CP-enhancement} is the chiral
perturbation theory \cite{vp1,gs}. In applying this method
including one-loop contributions  one should however take into
account that the region of validity of the naive quantum chiral
theory is only the low energy decomposition.

The first  paper on the meson form factors in chiral perturbation
theory \cite{fpp} revealed
 the resonance nature of this chiral decomposition.
 Taking into account the summation of the chiral series as
  the
 meson form factors in the investigation of the processes
 directly connected with the $K \to \pi$ transition (like
 nonleptonic kaon decays and  $K^+
 \to \pi^+e^{+}e^{-}(\mu^{+}\mu^{-},\nu\bar{\nu})$)
  %$K^+\pi^-\to e^{+}e^{-}(\mu^{+}\mu^{-}, \nu\bar{\nu})$
 is the main
 difference of the effective
  chiral Lagrangian approach applied in the present paper
  from other ones cited above.

 The introduction of the meson form factors in the chiral dynamics
 supposes that the low-energy dynamics can be separated
 by the Dirac formulation of the  Standard Model
 \cite{dir,pol,pvn6,211}, where all vector fields ($\gamma, W, Z$) contain
  instantaneous  potentials in the
 Lorentz frames defined
 by the {\it in} and {\it out} physical states
   considered in S- matrix as the irreducible
 representations of the Poincar\'e group.
 The dominance of the instantaneous
 interactions in the kaon-pion transitions
  differs our approach from the previous
 ones based on the heuristic Faddeev -- Popov approach \cite{fp1,f1},
 where instantaneous interactions were neglected.

 We suppose also that the quark content of the
  mesons determines hadronization
 of QCD \cite{5,6}  conserving its
 chiral and gauge symmetries.

 These differences allow us to study the possibility
 of establishing connection of the corresponding decay
probabilities with the  $\pi$ and $K$ meson form factors and
extracting information about form factors of $\pi $ and $K$ mesons
from the $K^+ \to \pi^+$ transition in the $K^+ \to
\pi^+e^{+}e^{-}(\mu^{+}\mu^{-},\nu\bar{\nu})$ and $K^+ \pi^-\to
e^{+}e^{-}(\mu^{+}\mu^{-},\nu\bar{\nu})$ processes. We present the
explicit forms of the amplitudes of these processes in terms of
meson  form factors  in the Standard Model of electroweak (EW)
interactions.

\section{Chiral Bosonization of EW Interaction}

 We begin with the Lagrangian of EW interaction of quarks
 \be\label{ch111}%\label{ew}
 \mathcal{L}_{(J)}=-\frac{e}{2\sqrt{2}\sin\theta_{W}}(J^{-}_{\mu}W^{+}_{\mu}
 +J^{+}_{\mu}W^{-}_{\mu}) ,
 \ee
 where
 $J^{+}_{\mu}=\bar{d}'\gamma_{\mu}(1-\gamma_{5})u; \quad \bar{d}'=d
 \cos\theta_{C}+s\sin\theta_{C}, \theta_{C}$ is a Cabbibo angle
 $\sin\theta_{C}=0.223$.
 The quark content of $\pi^{+}$ and $K^{+}$
 mesons $\pi^{+}=(\bar{d},u), K^{+}=(\bar{s},u),
  \overline{K}^{0}=(\bar{s},d) $  leads to the
 effective chiral hadron currents $J^{\pm}_\mu$ in
 the  Lagrangian (\ref{ch111})
 \be\label{chl1}
 J^{\pm}_{\mu}=[J^1_{\mu}{\pm}iJ^2_{\mu}]\cos\theta_{C}\,+
 [J^4_{\mu}{\pm}iJ^5_{\mu}]\,\sin\theta_{C}\,,
 \ee
 where using the Gell-Mann matrixes $\lambda^k$ one can define
 the meson current as \cite{vp1}
 \be\label{c3}
 i\sum\limits \lambda^k
 J^k_{\mu}=i\lambda^k(V^{k}_{\mu}-A^{k}_{\mu})^{k}=F^2_\pi
 e^{i\xi}\partial_\mu e^{-i\xi},
 \ee
\be\label{c4}
 \xi=F_\pi^{-1}\sum\limits_{k=1}^{8}M^k\lambda^k=F_\pi^{-1}\left(%
\begin{array}{ccc}
  \pi^0+\dfrac{\eta}{\sqrt{3}} & \pi^+\sqrt{2} & K^+\sqrt{2} \\
  \pi^-\sqrt{2}  & -\pi^0+\dfrac{\eta}{\sqrt{3}} & K^0\sqrt{2} \\
  K^-\sqrt{2} & \overline{K}^0\sqrt{2} & -\dfrac{2\eta}{\sqrt{3}} \\
\end{array}%
\right).
 \ee
 In the first orders in mesons one can write

 \be \label{chl2}
 V^{-}_{\mu}=\sqrt{2}\,\,[\,\sin\theta_{C}\,
 (K^{-}\partial_{\mu}\pi^{0}-\pi^{0}\partial_{\mu}K^{-} )\,
 +\cos\theta_{C}\,(\pi^{-}\partial_{\mu}\pi^{0}-\pi^{0}\partial_{\mu}\pi^{-})\,]
  \ee
  and
  \be\label{chl3}
  A^{-}_{\mu}=\sqrt{2}\,F_{\pi}\,(
 \partial^{\mu}K^{-}\sin\theta_{C} +
 \partial^{\mu}\pi^-\cos\theta_{C});
 \ee
 here $F_{\pi}\simeq  92$ MeV.
The right form of chiral Lagrangian of the electromagnetic
interaction of mesons  can be constructed by the covariant
derivative
 $
 \partial_{\mu}\chi^{\pm}\to D_{\mu}\chi^{\pm}
\equiv(\partial_{\mu}\pm ieA_{\mu})\chi^{\pm},
 $
 where $\chi^{\pm}=K^{\pm},\pi^{\pm}$.

 The main idea is a consideration of the instantaneous
 interactions in this model
 (that is associated with the Yukawa potential of massive vector bosons).

\section{Instantaneous Coulomb Interaction in QED}

 In order to explain the main idea of describing of massive bosons,
 let us consider  the action of photons $S_{\rm QED}=\int
 d^4x{\cal{L}}$,
 where
 \be\label{a2}
 {\cal{L}} = - \frac{1}{4}F_{\mu\nu} ~F_{\mu\nu} +j_{\mu} A_{\mu}
 - \bar {\psi} ( i\gamma \partial + m ) \psi
 ~~~~~~( \gamma \partial=\gamma_{\mu} \partial_{\mu}),
 \ee
 is the QED Lagrangian,  $A_{\mu}$ is a vector potential, $F_{\mu\nu} =
 \partial_{\mu}A_{\nu} - \partial_{\nu}A_{\mu}$, index
 $\mu,\nu= 0,1,2,3$, $\psi$ is the
 electron-positron field described by the Dirac bispinor,
 $j_{\mu}=e  \bar {\psi}  \gamma_{\mu} \psi$ is the charge current,
 and $e$ is the electron charge. %Lagrangian (\ref{a2}) results in the following
% equations
% \be\label{a5}
%      \partial_{\lambda}  ~F_{\lambda\nu}\equiv \square A_\nu
% -\partial_\nu (\partial_\mu A_\mu)  =-j_{\nu}~,
% \ee
% \be\nonumber%\label{a6}
% (i \gamma \partial +m)\psi~=-e \gamma_{\mu} \psi A_{\mu}~.
% \ee
% where $\square =[\partial^2_0-\partial^2_j]$
% is d'alambertian.

 %The acceptable photon propagator is used as the Lorentz one
%\begin{equation}\nonumber%\label{nonlpr}
% D(q)^{\gamma,(\rm lor)} = -\frac{\delta_{\mu
%\nu}}{q^2+i\varepsilon}\left[1- \frac{q_\mu
%q_\nu}{q^2+i\varepsilon}\right]~.
% \end{equation}
% A light-cone pole of this Lorentz frame free propagator
% does not reflect

 The instantaneous Coulomb interaction
 is an attribute
 of the Hamiltonian formulation in a concrete Lorentz frame in QED,
 because  the
 time component of the vector field $A_0$
 has the zero canonical
 momentum\footnote{Here $A_0=A_\mu n_{\mu}$ and
 $n_{\mu}$ is a unit timelike vector $n^2_{\mu}=1$
 that distinguishes timelike and spacelike
 components
 $
 [A_{\mu}]_{\rm n}=[(A_\mu n_{\mu}),~A_\nu -
 n_{\nu}(A_\mu n_{\mu})]_{n_\mu=(1,0,0,0)}=[A_0,A_k]
 $.}
 $ %\be\label{cm1}
 P_0={\partial {\cal L}}/{\partial (\partial_0A_0)}=0
 $. %\ee
 Therefore, Dirac \cite{dir}  supposed to consider $A_0$ as a classical field
  and excluded it  by the manifest resolution
 of the Gauss constraint
  $%\be\label{c1}
 \Delta A_0-\partial_0\partial_{k}A_k=-j_{0}
  $ %\ee
 (where $\Delta=\partial^2_{j}$)  and the gauge
 transformation
  \bea
 \label{r2v}
 A^{(\rm rad)}_0(A)&=&A_0+\partial_0\lambda^{(\rm rad)}(A)=A_0
 -\partial_0\frac{1}{\Delta} \partial_{k}~A_{k},\\
 \label{r2v1}
 A^{(\rm rad)}_l(A)&=&A_l+\partial_l\lambda^{(\rm rad)}(A)=
 A_l-\partial_l\frac{1}{\Delta} \partial_{k}~A_{k},\\
 \label{r3s}
 \psi^{(\rm rad)}(\psi,A)&=&e^{\imath e\lambda^{(\rm rad)}(A)}\psi,
 \eea
where
  $\lambda^{(\rm rad)}(A)=-\frac{1}{\Delta} \partial_{k}~A_{k}\equiv
  \frac{1}{4\pi}\int d^3y
  \frac{\partial_{k}~A_{k}(x_0,y_k)}{|x-y|}$.
 These gauge-invariant functionals were called the ``dressed fields'', or
 the ``radiation variables``.

 In terms of the gauge-invariant radiation variables the initial action of QED
 takes a form
\bea\label{1t2}
 S^{(\rm rad)}_{\rm QED}=
 \int d^4x \left[\frac{1}{2}(\partial_{\mu}A^{(\rm rad)}_k)^2
 +\frac{1}{2}
 j_0^{(\rm rad)}\frac{1}{\triangle}j_0^{(\rm rad)} - A^{(\rm rad)}_kj^{(\rm rad)}_{k}-
 \bar \psi^{(\rm rad)}(i\gamma\partial+M)\psi^{(\rm rad)}\right].
 \eea
 This Lagrangian contains
 the Coulomb
  instantaneous interaction forming atomic and molecular bound states
  in the lowest order of the ``radiation corrections'' $A_k^{(\rm
  rad)}=0$.
  The radiation action (\ref{1t2}) corresponds to the propagator
 \bea\label{wr}
  D^{\gamma,{(\rm rad)}}_{\mu\nu}(q) = \delta_{\mu
 0}\delta_{\nu 0}\frac{1}{\vec{q}^2} &+&\delta_{\mu i}\delta_{\nu j}
 \left(\delta_{ij}-\frac{q_iq_j}{\vec{q}^2}\right)\frac{1}{q^2+i\varepsilon}
 \equiv\\
\label{1wr}
 &\equiv&-\frac{\delta_{\mu \nu}}{{q}^2+i\varepsilon} +
 \frac{(q_0\delta_{\mu 0})(q_0\delta_{\nu 0})
 -(q_i\delta_{\mu i})(q_j\delta_{\nu j})}{\vec{q}^2[q^2+i\varepsilon]}.
\eea
 Recall that Faddeev \cite{f1} considered this Dirac approach
 (\ref{1t2}) to QED
 applied by Schwinger \cite{sch2} for non-Abelian theory
 as the
 foundation of the Faddeev -- Popov  (FP)
 heuristic approach \cite{fp1}, where
  the instantaneous interactions disappear. Really,
 one can prove  \cite{f1} that the last (longitudinal) term
  in Eq. (\ref{1wr}) can
 be neglected  for the class of the elementary particle scattering
 amplitudes, where charge currents are conserving and satisfy the equality
 $j_0q_0-j_kq_k=0$.
 However, in the case of calculation of  the bound state spectrum
  the last term cannot
 be neglected, because  it
 gives in (\ref{wr}) the instantaneous singularity forming observable atoms.

 Using the  Lorentz transformations
 (proved at the level of Poincare algebra of observables
 \cite{z})
 one can find the
 photon propagator in an arbitrary frame as follows \cite{pol},
 \begin{equation}
D^{\gamma,{(\rm
rad)}}_{\mu\nu}(q|n)=\frac{n_{\mu}n_{\nu}}{|{q^{\perp}}^2|}
-\left(\delta^{\perp}_{\mu\nu}-\frac{q^{\perp}_{\mu}q^{\perp}_{\nu}}
{|{q^{\perp}}^2|}\right)\frac{1}{q^{2}+i\epsilon} \label{gr},
\end{equation}
where
 $q^{\perp}_{\mu}=q_{\mu}-n_{\mu}(qn), ~~
 \delta^{\perp}_{\mu\nu}=\delta_{\mu\nu}-n_{\mu}n_{\nu},
 $ and
 $n_{\mu}$ is the timelike vector reflecting the initial conditions defined
 by physical states of the quantized fields \cite{211,5,6}.

\section{Instantaneous  Interaction in SM}

  SM was formulated \cite{db} on the basis of the FP heuristic approach
  \cite{fp1} without instantaneous interactions. Therefore, we
  shall use here the
 generalization of the fundamental radiation variables in QED (\ref{gr})
to massive vector fields,  described by the action,
 \begin{equation}\nonumber
  W=\int
d^4x\left[-\frac{1}{2}(\partial_\mu W_{\nu}^{+}-\partial_\nu
W_{\mu}^{+})(\partial_\mu W_{\nu}^{-}-\partial_\nu
W_{\mu}^{-})+M^{2}_{W}W_{\mu}^{+}W_{\mu}^{-}+J^{-}_\mu
W_\mu^++J^{+}_\mu W_\mu^-\right] \label{Lem}
\end{equation}
that considered  in \cite{pvn6} using the manifest resolution of
the Gauss equation and  ``dressing''  all fields by
 the gauge transformation
 with $\lambda^{\pm(\rm rad)}=[{1}/{(M_W^2-\triangle)}]\partial_kW^{\pm}_k$.

The action of W-bosons in SM in terms of the radiation variables
 contains the instantaneous Yukawa interaction of
 the currents (\ref{chl2}) and (\ref{chl3})
 with the Hamiltonian
 \be\label{yk}
 H_{\rm Yuk}=\frac{G_{F}}{\sqrt{2}}\int
 d^3x J^-_0\frac{1}{M_{W}^2-\triangle}J^+_0~, \qquad {\rm where}\qquad
 \frac{G_{F}}{\sqrt{2}}=\frac{e^2}{8\sin^2\theta_W}.
 \ee

 The radiation variables lead to a propagator of
type of the  QED one (\ref{wr})
\begin{equation}
\label{wr1} D^{W,{(\rm rad)}}_{\mu\nu}(q) = \delta_{\mu
0}\delta_{\nu 0}\frac{1}{\vec{q}^2+M_{W}^2} +\delta_{\mu
i}\delta_{\nu j} \left(\delta_{ij}-\frac{q_iq_j}
{\vec{q}^2+M_{W}^2}\right)\frac{1}{q^2-M_{W}^2+i\varepsilon}~.
\end{equation}
  which has
 no singularity in the
massless limit and is well behaved for large momenta. As it was
shown in  \cite{pvn6}, the Lorentz transformations of classical
radiation variables coincide with the  quantum ones  and they both
(quantum and classical) correspond to the transition to another
Lorentz reference frame distinguished by another time-axis, where
the propagator takes the form
\begin{equation}
D^{W,{(\rm
rad)}}_{\mu\nu}(q|n)=\frac{n_{\mu}n_{\nu}}{M_W^2+|{q^{\perp}}^2|}
-\left(\delta^{\perp}_{\mu\nu}-\frac{q^{\perp}_{\mu}q^{\perp}_{\nu}}
{M_W^2+|{q^{\perp}}^2|}\right)\frac{1}{q^{2}-M_W^2+i\epsilon}
\label{gv2}
\end{equation}
with
 $
 q^{\perp}_{\mu}=q_{\mu}-n_{\mu}(qn),~~~
 \delta^{\perp}_{\mu\nu}=\delta_{\mu\nu}-n_{\mu}n_{\nu}
 $.
 %The time-axis $n_\mu$ is given by the in (out) physical states \cite{5,6}.
%Further we would like to apply  this propagator to the description
%of  physical processes with axial and vector currents.

\section{The $K \to \pi$ Transition Amplitude and the Rule $\Delta T=\dfrac12$}

 One can estimate
 the $K^+\to \pi^+$ transition amplitude
 \be\label{kp1}
 <\pi^+|-i\int dx^0 H_{\rm Yuk}|K^+>=i(2\pi)^{4}\delta^{4}(k-p)G_{\rm EW}\Sigma(p^{2}),
 \ee
where
 \be\label{g}
  G_{\rm EW}= \frac{\sin\theta_{C} \cos\theta_{C}}{8
  M^{2}_{W}}\frac{e^{2}}{\sin^{2}\theta_{W}}\equiv
  \sin\theta_{C} \cos\theta_{C}\frac{G_F}{\sqrt{2}}
 \ee
 is the constant.
   Making the normal ordering of the $\pi^0$
 fields
 \be\label{n0}
 \triangle(\vec x-\vec y)=<0|\pi^i(x)\pi^{i'}(y)|0>=
  \delta^{ii'}\int \frac{d^3l}{(2\pi)^3}
   \frac{e^{i\vec l\cdot (\vec x -\vec y)}}{2E_\pi(\vec l)},
  \ee
  where $E_{\pi}(\vec{l})=\sqrt{m^{2}_{\pi}+\vec{l}^{2}}$ is the
 energy of $\pi$-meson, in the  product of vector currents,
  %in the rest frame $p_\mu=(p_0,0,0,0)$,
  one can find the
 expression in the lowest order of chiral perturbation theory
 \be\label{kp2}
  \Sigma(p^{2})=2p_0^2\left[\frac{F^2_{\pi}}{M_{W}^{2}}+
  \frac{1}{(2\pi)^3}\int\frac{d^3l}{2E_\pi(\vec{l})}
  \frac%{f_{K\pi W}(\vec{l}^2)f_{\pi\pi W}(\vec{l}^2)}
  {1}{M^2_W+\vec{l}^2}\right]
  \equiv
  2p_0^2\frac{F^2_{\pi}}{M_{W}^{2}}g_8;
  \ee
  %$f^{V}_{K\pi W}(\vec{l^{2}})$ and
% $f^{V}_{\pi \pi W}(\vec{l^{2}})$ are the weak vector form factors
% introduced   as the natural regularization
% of meson  instantaneous interactions by the next orders \cite{fpp},
 here
\be\label{vv3}
 g_8=1+\frac{M^{2}_{W}}{F^2_{\pi}(2\pi)^{3}}
 \int\frac{d^3 l}{2E_{\pi}(\vec{l})}
 \frac
 %{f^{V}_{K\pi W}(\vec{l^{2}})f^{V}_{\pi \pi W}(\vec{l}^{2})}
 {1}{M^{2}_{W}+\vec{l}^{2}}%+...
 %=1+\frac{\triangle(0)}{F^2_{\pi}}+...,
 \ee is
 the parameter of   the enhancement of the probability
 of $K^+ \to \pi^+$ transition.
 In the low-energy chiral perturbation theory  \cite{fpp},
  the summation of the chiral series can be parameterized by
  the meson form factors $f^{V}_{K\pi W}(-\vec{l^{2}})$ and
 $f^{V}_{\pi \pi W}(-\vec{l^{2}})$
 \cite{bvp,belkov01,CP-enhancement}.

  Finally, the instantaneous interaction can
 give the  result\be\label{1vv3}
 g_8=1+\frac{M^{2}_{W}}{F^2_{\pi}(2\pi)^{3}}\int\frac{d^3 l}
 {2E_{\pi}(\vec{l})}\frac{f^{V}_{K\pi
W}(\vec{-l^{2}})f^{V}_{\pi
 \pi W}(-\vec{l}^{2})}{M^{2}_{W}+\vec{l}^{2}},
 \ee
 where $g_8=5.1$ \cite{ecker}. The relation (\ref{1vv3}),
  can be considered as the low-energy sum rule
 for meson form factors  in the space-like region.

 This result shows that the simultaneous vector interaction
 can increase the axial interaction $K^+\to \pi^+$ transition
 in $g_8$ times and give a new term describing the
  $K^0\to \pi^0$ transition proportional
 to $g_8-1$.% accepted now for description of
%  the observational data \cite{7,b}.

 Using
  the covariant perturbation theory \cite{pv}\footnote{The
 covariant perturbation theory was developed as the
 series $J_0^k(\gamma
 \oplus\xi)=J_0^k(\xi)+ F^2_\pi
 \partial_0 \gamma^k +\gamma^if_{ijk}J_0^j(\xi)+O(\gamma^2)$
 with respect to quantum fields $\gamma$ added to $\xi$
 as  the product $e^{i\gamma}e^{i\xi}\equiv e^{i(\gamma \oplus\xi)}$
  \cite{pv}.} and relativistic invariance ($J^2_0\to J^2_\mu$)
 one can  calculate this enhancement in terms of currents
 as the effective  Lagrangians  \cite{kp}
 \be \mathcal{L}_{(\Delta T=\frac{1}{2})}=
\frac{G_{F}}{\sqrt{2}}g_{8}\cos\theta_{C}\sin\theta_{C}
\Big[(J^1_{\mu}+iJ^2_{\mu})(J^4_{\mu}-iJ^5_{\mu})-
(J^3_{\mu}+\frac{1}{\sqrt{3}}J^8_{\mu})(J^6_{\mu}-iJ^7_{\mu})+h.c.\Big],
\ee \be \mathcal{L}_{(\Delta T=\frac{3}{2})}=
\frac{G_{F}}{\sqrt{2}}\cos\theta_{C}\sin\theta_{C}
\Big[(J^3_{\mu}+\frac{1}{\sqrt{3}}J^8_{\mu})(J^6_{\mu}-iJ^7_{\mu})+h.c.\Big]
\ee  in satisfactory agreement with experimental data within the
 accuracy  $20\div 30\%$ \cite{vp1,kp}.

\section{The  $K^+\to\pi^++l+\bar l$ amplitude}

\begin{figure}
\centering
\begin{minipage}[c]{0.45\hsize}
\begin{picture}(200,100)(0,0)\Vertex(150,50){5}\Vertex(50,50){5}\Vertex(100,50){5}
\ZigZag(50,50)(100,50){5}{5}\ArrowLine(100,50)(150,50)\ArrowLine(150,50)(200,50)
\Photon(150,50)(150,100){5}{3} \SetWidth{2.0}
\ArrowLine(0,50)(50,50)
\Text(5,90)[]{$(a)$}\Text(25,30)[]{$K^{+}(k)$}
\Text(125,30)[]{$\pi^{+}(k)$}\Text(175,30)[]{$\pi^{+}(p)$}\Text(75,30)[]{$W^{+}(k)$}
\Text(170,85)[]{$\gamma^{*}(q)$}
\end{picture}
\end{minipage}
\hspace*{5mm}
\begin{minipage}[c]{0.45\hsize}
\begin{picture}(200,100)(0,0)\Vertex(150,50){5}\Vertex(50,50){5}\Vertex(100,50){5}
\ZigZag(100,50)(150,50){5}{5}\ArrowLine(150,50)(200,50)
\Photon(50,50)(50,100){5}{3} \SetWidth{2.0}
\ArrowLine(0,50)(50,50)\ArrowLine(50,50)(100,50)\Text(5,90)[]{$(b)$}
\Text(25,30)[]{$K^{+}(k)$}
\Text(125,30)[]{$W^{+}(p)$}\Text(175,30)[]{$\pi^{+}(p)$}\Text(75,30)[]{$K^{+}(p)$}
\Text(70,85)[]{$\gamma^{*}(q)$}
\end{picture}
\end{minipage}
\begin{minipage}[c]{0.45\hsize}
\begin{picture}(200,100)(0,0)
\Vertex(70,50){5}\Vertex(130,50){5}
\ZigZag(70,50)(130,50){5}{5}\ArrowLine(130,50)(200,50)
\Photon(130,50)(130,100){5}{3} \SetWidth{2.0}
\ArrowLine(0,50)(70,50)
\Text(5,90)[]{$(c)$}\Text(35,30)[]{$K^{+}(k)$}
\Text(100,30)[]{$W^{+}(k)$}\Text(165,30)[]{$\pi^{+}(p)$}
\Text(150,85)[]{$\gamma^{*}(q)$}
\end{picture}
\end{minipage}\hspace*{5mm}
\begin{minipage}[c]{0.45\hsize}
\begin{picture}(200,100)(0,0)
\Vertex(70,50){5}\Vertex(130,50){5}
\ZigZag(70,50)(130,50){5}{5}\ArrowLine(130,50)(200,50)
\Photon(70,50)(70,100){5}{3} \SetWidth{2.0}
\ArrowLine(0,50)(70,50)
\Text(5,90)[]{$(d)$}\Text(35,30)[]{$K^{+}(k)$}
\Text(100,30)[]{$W^{+}(p)$}\Text(165,30)[]{$\pi^{+}(p)$}
\Text(90,85)[]{$\gamma^{*}(q)$}
\end{picture}
\end{minipage}
\caption{Axial current contribution} \label{ac}\end{figure}
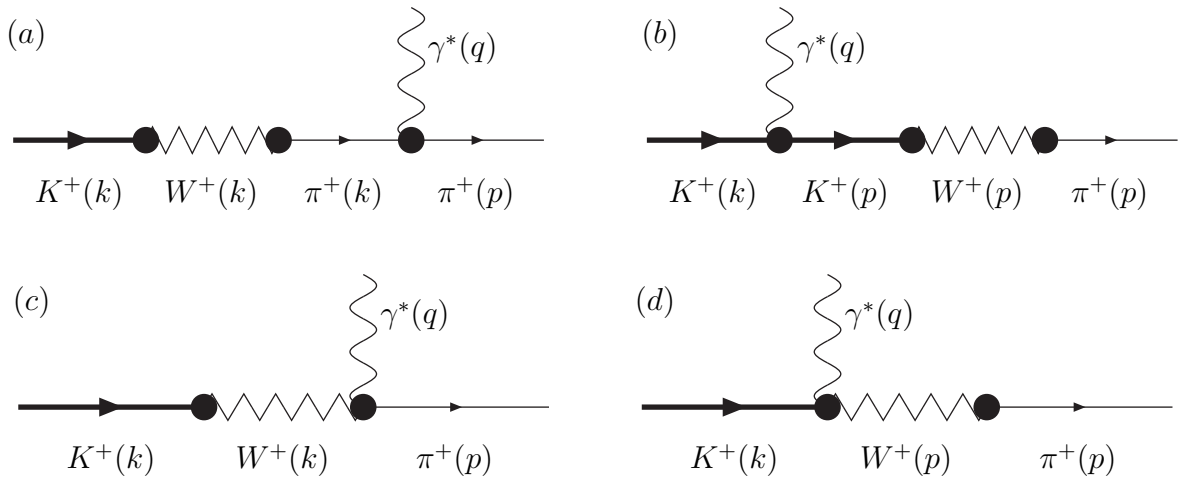

 The result of calculation of the axial contributions to the
amplitude of the process $K^+\to\pi^++l+\bar l$ within the
framework of chiral Lagrangian (\ref{ch111}), (\ref{c3}) including
phenomenological meson form factors shown  in Fig.
 \ref{ac} as a fat dot  takes the form

%\begin{figure}[t]
%\vspace{-1cm}
% \begin{center}
%\includegraphics[width=0.6\textwidth,clip]{1s.eps}
%\caption{{text} \label{fig2}}
%\end{center}
%\end{figure}

 \be\label{ampl}%\label{aatp}
  T^{AA}(K^+\to \pi+l^+l^-)= 2eG_{\rm EW}L_{\nu}
 D^{\gamma(rad)}_{\mu\nu}(q)(k_{\mu }+p_{\mu })\,\,\,
 t^{AA}(q^2,k^2,p^2),
 \ee
 where $G_{\rm EW}$ is the constant (\ref{g}),
  $L_{\mu}=\bar{l}\gamma_{\mu}l$ is leptonic current,
 \be\label{tp}
 t^{AA}(q^2,k^2,p^2)=F^2_{\pi}
 \left[\frac{f^{V}_{\pi}(q^{2})k^{2}}{m^{2}_{\pi}-k^{2}-
 i\epsilon}+\frac{f^{V}_{K}(q^{2})p^{2}}{M^{2}_{K}-p^{2}-i\epsilon}+
 \frac{f^{A}_{K}(q^{2})+f^{A}_{\pi}(q^{2})}{2}\right],
 \ee
 and
 $f^{(A,V)}_{\pi,K}(q^{2})$ are meson form factors.
 In this expression the propagator
 $D_{\mu\nu}^{W,{(\rm rad)}}(p|n)$ given by Eq. (\ref{gv2}) keeps only the Yukawa potential
 part $\dfrac{p_\mu p_\nu}{p^2 M^2_W}$.

 On the mass-shell the sum (\ref{tp})  takes the form
 \bea\label{t0}
 t^{AA}(q^2,M_K^2,m_\pi^2)&=&\\\nonumber
 = t(q^2)&=&F^2_\pi  \Bigg[\frac{f^{A}_{K}(q^{2})+f^{A}_{\pi}(q^{2})}{2}
 - f^{V}_{\pi}(q^{2})+
 [f^{V}_{K}(q^{2})- f^{V}_{\pi}(q^{2})]\frac{m_\pi^2}{M_K^2-m_\pi^2}
  \Bigg]~.
 \eea

As can be seen in \cite{bvp,ecker}, the amplitude for $K^{+}\to
\pi^{+}+l+\bar l $ vanishes at the tree level, where  form factors
are equal to unit
  $%\be\label{tpv}
t(q^2)|_{f^{V}=f^{A}=1}=0.
 $ %\ee
 It is well known that these form factors in the limit $q^2\to 0$
  take the form
  $%\be
 f^{A,V}_{\pi,K}(q^{2})=1-\frac{q^{2}}{6}<r^{2}>^{(A,V)}_{\pi,K}
 $, %\ee
  where  $<r^{2}>^{(A,V)}_{\pi,K}$ are the axial and vector
 mean square radii of mesons, respectively, in agreement with
 the chiral perturbation theory \cite{bvp,vp1,fpp}.

%\hspace{0.5cm}

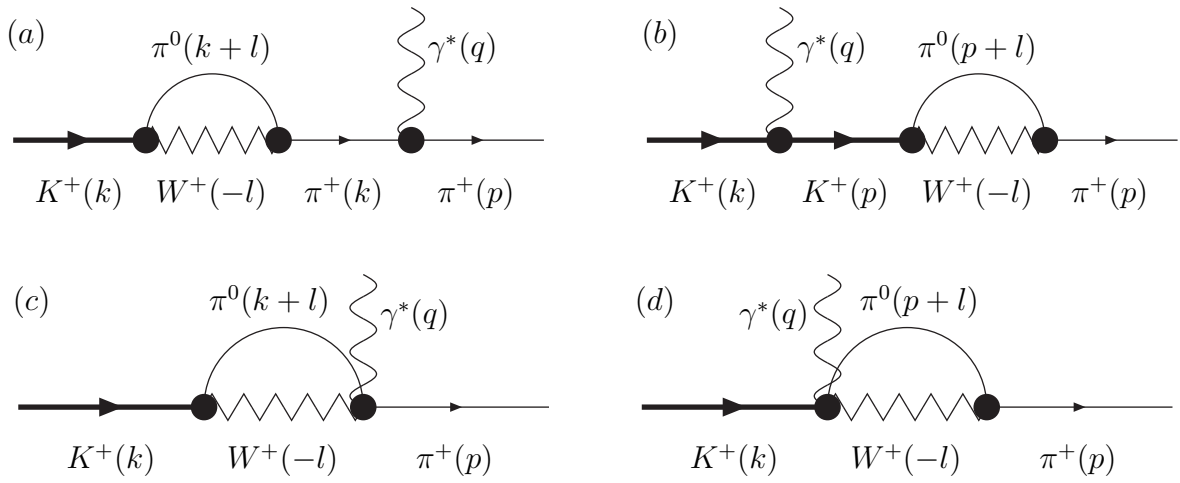
\begin{figure}
\centering
\begin{minipage}[c]{0.45\hsize}
\begin{picture}(200,100)(0,0)\Vertex(150,50){5}\Vertex(50,50){5}\Vertex(100,50){5}\CArc(75,50)(25,0,180)
\ZigZag(50,50)(100,50){5}{5}\ArrowLine(100,50)(150,50)\ArrowLine(150,50)(200,50)
\Photon(150,50)(150,100){5}{3} \SetWidth{2.0}
\ArrowLine(0,50)(50,50) \Text(5,90)[]{$(a)$}
\Text(25,30)[]{$K^{+}(k)$}
\Text(125,30)[]{$\pi^{+}(k)$}\Text(75,85)[]{$\pi^{0}(k+l)$}\Text(175,30)[]{$\pi^{+}(p)$}\Text(75,30)[]{$W^{+}(-l)$}
\Text(170,85)[]{$\gamma^{*}(q)$}
\end{picture}
\end{minipage}
\hspace*{5mm}
\begin{minipage}[c]{0.45\hsize}
\begin{picture}(200,100)(0,0)\Vertex(150,50){5}\Vertex(50,50){5}\Vertex(100,50){5}\CArc(125,50)(25,0,180)
\ZigZag(100,50)(150,50){5}{5}\ArrowLine(150,50)(200,50)
\Photon(50,50)(50,100){5}{3} \SetWidth{2.0}
\ArrowLine(0,50)(50,50)\ArrowLine(50,50)(100,50)\Text(5,90)[]{$(b)$}
\Text(25,30)[]{$K^{+}(k)$}\Text(125,85)[]{$\pi^{0}(p+l)$}
\Text(125,30)[]{$W^{+}(-l)$}\Text(175,30)[]{$\pi^{+}(p)$}\Text(75,30)[]{$K^{+}(p)$}
\Text(70,85)[]{$\gamma^{*}(q)$}
\end{picture}
\end{minipage}
\begin{minipage}[c]{0.45\hsize}
\begin{picture}(200,100)(0,0)
\Vertex(70,50){5}\Vertex(130,50){5}\CArc(100,50)(30,0,180)
\ZigZag(70,50)(130,50){5}{5}\ArrowLine(130,50)(200,50)
\Photon(130,50)(130,100){5}{3} \SetWidth{2.0}
\ArrowLine(0,50)(70,50)
\Text(5,90)[]{$(c)$}\Text(35,30)[]{$K^{+}(k)$}\Text(95,90)[]{$\pi^{0}(k+l)$}
\Text(100,30)[]{$W^{+}(-l)$}\Text(165,30)[]{$\pi^{+}(p)$}
\Text(150,85)[]{$\gamma^{*}(q)$}
\end{picture}
\end{minipage}\hspace*{5mm}
\begin{minipage}[c]{0.45\hsize}
\begin{picture}(200,100)(0,0)
\Vertex(70,50){5}\Vertex(130,50){5}\CArc(100,50)(30,0,180)
\ZigZag(70,50)(130,50){5}{5}\ArrowLine(130,50)(200,50)
\Photon(70,50)(70,100){5}{3} \SetWidth{2.0}
\ArrowLine(0,50)(70,50)\Text(5,90)[]{$(d)$}
\Text(35,30)[]{$K^{+}(k)$}\Text(105,90)[]{$\pi^{0}(p+l)$}
\Text(100,30)[]{$W^{+}(-l)$}\Text(165,30)[]{$\pi^{+}(p)$}
\Text(50,85)[]{$\gamma^{*}(q)$}
\end{picture}
\end{minipage}
\caption{Vector current contribution} \label{vc}\end{figure}

 The result of calculation of vector contributions to the amplitude
 of the process $K^+\to\pi^++l+\bar l$ within
 the framework of chiral Lagrangian (\ref{ch111}), (\ref{c3})
 including phenomenological meson form factors  shown  in Fig.
 \ref{vc} as a fat dot  takes the form
 \be T^{VV{(\rm rad)}}(K^+\to \pi+l^+l^-)=
2eG_{\rm EW}L_{\nu} D^{\gamma(rad)}_{\mu\nu}(q)(k_{\mu } +p_{\mu
})\,\,\, t^{VV}(q^2,k^2,p^2), \label{ampl1}
 \ee
 where

 \begin{eqnarray}
 \nonumber
  t^{VV}(q^2,k^2,p^2)&=&\frac{1}{2(2\pi)^{2}}\int\frac{d|\vec l| ~~
\vec{l^{2}}}{E_{\pi}(\vec{l})} \Bigg\{{f^{V}_{K\pi
W}(\vec{l^{2}})f^{V}_{\pi
 \pi W}(\vec{l}^{2})}\Bigg[
\frac{f^V_{\pi}(q^2)k^{2}} {m^{2}_{\pi}-k^{2}-i\epsilon} +
 \\\nonumber
 +\frac{f^V_{K}(q^2)p^{2}}{M^{2}_{K}-p^{2}-i\epsilon}\Bigg]&+&
   \frac{f_\pi^A(q^2)f^V_{K \pi W}(\vec{l^{2}})
   f_{\pi \pi W\gamma}^{V}(\vec{l^{2}})+
   f_K^A(q^2)f^V_{\pi \pi W}(\vec{l^{2}})
   f_{K \pi W\gamma}^{V}(\vec{l^{2}})}{2}
   \Bigg\};
\end{eqnarray}
 here $f_{\pi \pi W\gamma}^{V}(\vec{l^{2}}),~f_{K \pi
W\gamma}^{V}(\vec{l^{2}})$ are the four particle interaction form
factors which should coincide with the three particle interaction
 form factors $f_{\pi \pi W\gamma}^{V}=f_{\pi \pi W}^{V},~
 f_{K \pi W\gamma}^{V}=f_{K \pi W}^{V}$ in accordance with
 the gauge invariance of the strong interactions.

 Using the result (\ref{vv3}) one can write
 on the mass shell $k^{2}=M_{K}^{2}, p^{2}=m_{\pi}^{2}$:
\bea\label{tvv}
t^{VV}(q^{2},M_{K}^{2},m_{\pi}^{2})&=&(g_8-1)t(q^{2}),\\
t^{AA}(q^{2},M_{K}^{2},m_{\pi}^{2})+
t^{VV}(q^{2},M_{K}^{2},m_{\pi}^{2})&=&g_8\,\,t(q^{2}),
 \eea
 where $g_8,t(q^{2})$ are given by Eqs. (\ref{vv3}) and (\ref{t0}), respectively.

 Finally, the result of calculation of axial and vector contributions to
 the amplitude of the process $K^+\to\pi^++l+\bar l$ within the
 framework of chiral Lagrangian (\ref{chl1}),(\ref{chl2}) including
 phenomenological meson form factors takes the form
 \be\label{amplv}
  T^{\rm (rad)}_{(K^+\to \pi^+l^+l^-)}=2e g_8
  G_{\rm EW}   L_{\nu}
 D^{\gamma (rad)}_{\mu\nu}(q)(k_{\mu }+p_{\mu })\,\,t(q^2).
 \ee

 This amplitude leads to  the total decay rate  for the transition $K^+ \to
 \pi^+e^{+}e^{-}$
 \be\label{f2}
 \Gamma_{(K^+ \to \pi^+e^{+}e^{-})}((M_K-m_\pi)^2)=\overline{\Gamma}_{e^+e^-}
 \int\limits_{4m^2_e}^{(M_K-m_\pi)^2}
 {\frac{d q^2}{M_K^2} \rho(q^2)}F(q^2),
 %\left|\frac{t(q^2)}{q^2}\right|^2
 \ee
 where
 \bea\label{gf3}
 \overline{\Gamma}_{e^+e^-}=\frac{(s_1c_1c_3)^2g_8^2G^2_F}{(4\pi)^4}
 \frac{\alpha^2 M_K^5}{24\pi}|_{g_8=5.1}=1.37 \times 10^{-22}
 \mbox{\rm GeV},
 \eea
 \bea\label{f3}\nonumber
 F(q^2)&=& \left[\frac{(4\pi)^2t(q^2)}{q^2}\right]^2=\\\label{1f3}
  &=&\left[\frac{(4\pi F_\pi)^2}{q^2}\right]^2
  \Bigg[\frac{f^{A}_{K}(q^{2})+f^{A}_{\pi}(q^{2})}{2}
 - f^{V}_{\pi}(q^{2})+
 [f^{V}_{K}(q^{2})-
 f^{V}_{\pi}(q^{2})]\frac{m_\pi^2}{M_K^2-m_\pi^2},
  \Bigg]^2
 \eea
and
 $$
 \rho(q^2)=\left(1-\frac{4m_l^2}{q^2}\right)^{1/2}\times\left(1+\frac{2m_l^2}{q^2}\right)~
 \lambda^{3/2}
 (1,q^2/M^2_K,m^2_\pi/M^2_K);
 $$
 here $\lambda (a,b,c)=a^2+b^2+c^2-2(ab+bc+ca)$,
  $s_1 c_1 c_3$ is the product of
  Cabibbo-Kobayashi-Maskawa matrix elements $V_{ud}V_{us}$.
 %with $c_1=\cos\theta_C, s_1c_3=\sin\theta_C$.

 The
 mechanism of enhancement $|\Delta T| = \frac{1}{2}$
 considered in this paper can be generalized to a description
 of other processes including $K^+\to \pi^+ \nu \bar{\nu}$
 by replacing the $\gamma$-propagator by the Z- boson one, so that
 we have the relations
 \bea\label{dr}\nonumber
 &&\left[\frac{q^2}{(4\pi F_\pi)^2}\right]^2\!\!\!\frac{M^2_K}{\rho(q^2)
\overline{\Gamma}_{e^+e^-}} \frac{d
 \Gamma_{(K^+ \to \pi^+e^{+}e^{-})}(q^2)}{
  \,\,\,\,d q^2}%\!\!\!\!\!\!\!\!
=\\\nonumber
  &&=\left[\frac{M_Z^2}{(4\pi F_\pi)^2}\right]^2\frac{M^2_K}{\rho(q^2)
\overline{\Gamma}_{\nu\bar \nu}} \frac{d
 \Gamma_{(K^+ \to \pi^+\nu \bar \nu)}(q^2)}{
  \,\,\,\,d q^2}=%\!\!\!\!\!
\\\label{dr1}
 &&=%\!\!\!\!\!\!\!
\Bigg[\frac{f^{A}_{K}(q^{2})\!+\!f^{A}_{\pi}(q^{2})}{2}
 - f^{V}_{\pi}(q^{2})+
 [f^{V}_{K}(q^{2})- f^{V}_{\pi}(q^{2})]\frac{m_\pi^2}{M_K^2-m_\pi^2}
  \Bigg]^2.
 \eea
Thus, exploiting
 the instantaneous weak interaction
 mechanism of enhancement in the
 $K \to \pi$ transition probability
 and QCD symmetry we derive the sum rule of EW vector
 meson form factors  given by Eqs. (\ref{vv3}), (\ref{t0}), (\ref{f2})
  and their relation to the differential $K \to \pi e^+ e^-$ decay
  rate (\ref{dr}).

\subsection{The form factor probe of
the differential $K \to \pi e^+ e^-$ decay
  rate}
 %NA48/2 CERN experiment

  The estimation of
 the  meson loop contribution was made
 in \cite{ecker} where
  a  function  $\hat \phi^2(q^2)$ was used
   instead of the
   form factor rate  $F(q^2)$ (\ref{1f3})%\cite{vp1,fpp}
  \bea\nonumber
   F(q^2)~~\Longrightarrow~~ \hat \phi^2(q^2)%|_{q^2=0}= 1
  \eea
  It was shown
  that the values $g_8=5.1$, $\hat \phi(0)=1$  gave
the total decay rate $\Gamma_{(K^+ \to
\pi^+e^{+}e^{-})}=1.91\times
 10^{-23}$ GeV
  in the satisfactory agreement with the experimental data
 $\Gamma_{(K^+ \to \pi^+e^{+}e^{-})}=1.44\pm 0.27\times
 10^{-23}$ GeV \cite{7,b}. However, the main contribution
 goes from the baryon loops  \cite{bvp,belkov01,CP-enhancement}.
 Therefore, we discuss here
 the value of the differential $K \to \pi e^+ e^-$ decay
  rate (\ref{1f3}) in the chiral perturbation theory \cite{vp1,fpp} where
 both the pion loop contribution $ \Pi_\pi$ and baryon ones
 lead to the meson form factors and resonances \cite{bvp,fpp}
 in the Pad\'e-type approximation
   %the value of the total decay rate(\ref{f2})
 \bea\label{1vform}
 f_{\pi}^{V}(q^2)&=&1+M^{-2}_\rho q^2+
 \alpha_{0}\Pi_\pi(q^2)+...\simeq\frac{1}{1-M^{-2}_\rho q^2-
 \alpha_{0}\Pi_\pi(q^2)}
 \\\label{2vform}
 f_{\pi}^{A}(q^2)&\simeq&f_{K}^{A}(q^2)=1+M^{-2}_a
 q^2+...\simeq\frac{1}{1-M^{-2}_a q^2}
 \eea
where
 \be\label{m1}\alpha_0=\dfrac{m_\pi^2}{(4\pi F_\pi)^2}\dfrac{4}{3}\simeq
 \dfrac{2}{103},
 \ee
\bea\label{M1}
 \Pi_\pi(t)= (1-\bar t)\left(\dfrac{1}{\bar t}-1\right)^{1/2}
 \arctan\left(\dfrac{\bar t^{1/2}}{(1-\bar t)^{1/2}}\right)-1;
 ~~~~~0<t<(2m_\pi)^2
 \\
 \Pi_\pi(t)= \dfrac{\bar t-1}{2}\left(1-\frac{1}{\bar t}\right)^{1/2}
\left\{i\pi
 -\log
 \dfrac{\bar t^{1/2}+(\bar t-1)^{1/2}}{\bar t^{1/2}-(\bar t-1)^{1/2}}
 \right\}+1;
 ~~~~~(2m_\pi)^2<t
 \eea
 is the pion loop contribution \cite{vp1,fpp}, and
 $M_\rho=771$ MeV and
  $M_a=980$ MeV
 are  the values of resonance masses \cite{7} that can be
 determined by the baryon loops  \cite{bvp,belkov01,CP-enhancement,fpp}.

 The constant approximation of the type of \cite{ecker}
 \be\label{M2}
 F(q^2=0)=F(0) =
  {(4\pi F_\pi)^4}\frac{[f_\pi^A(q^2)-f_\pi^V(q^2)]^2}{(q^2)^2}|_{q^2=0}=
  (4\pi F_\pi)^4[M_\rho^{-2}-M_a^{-2}]^2= 0.74
  \ee
  corresponds  to the value of $\Gamma_{(K^+ \to \pi^+e^{+}e^{-})}=1.41\times
 10^{-23}$ GeV
 in satisfactory agreement with
 the experimental data $\Gamma_{(K^+ \to \pi^+e^{+}e^{-})}=1.44\pm 0.27\times
 10^{-23}$ GeV \cite{7,b}.

 The form factors (\ref{1vform}) and (\ref{2vform}) below
 the two particle threshold $q^2<4m^2_\pi$ determines
  the differential rate $F(q^2)$
 as a function of $q^2$
 \be\label{f4}
 F(q^2)=\frac{F(0)}{[1-M_\rho^{-2}q^2]^2[1-M_a^{-2}q^2]^2}~~\longrightarrow~~
 \frac{d}{d q^2}\log F(q^2)|_{q^2=0}\simeq 2[M_\rho^{-2}+M_a^{-2}]\simeq 5
 \mbox{\rm GeV}^{-2}
 \ee
 At the region above
 the two particle threshold $4m^2_\pi<q^2<(M_K-m_\pi)^2$ there is
  a jump of the differential rate $F(q^2)$ at the level of
 $5\div10$ \%
\be\label{f5}
 F(q^2)=(4\pi F_\pi)^4\dfrac{[M_\rho^{-2}-M_a^{-2}]^2+
 {\alpha^2_0\pi^2(q^2-4m_\pi^2)}/{(4q^6)}}
 {[1-M_\rho^{-2}q^2]^2[1-M_a^{-2}q^2]^2}
 \ee

 These results (\ref{M2}), (\ref{f4}), and (\ref{f5}) are  arguments that the detailed
  investigation of the differential $K \to \pi e^+ e^-$ decay
  rate in the NA48/2 CERN experiment
 can give us information about  the meson form factors.
 To make much more
 realistic analysis,
 we have to use the unitary and analytic
 model for meson form factors \cite{duby} describing very well all
 experimental data.

\section*{Conclusion}

 We have seen that experimental data testify to that
  the kaon-pion transition can be
     obtained, in SM model, by the normal ordering
 of the pion field in the instantaneous  interaction (\ref{yk})
  as one of the indispensable attributes of
 the Dirac  approach  to SM \cite{211}.
 This  interaction can explain the rule $\triangle T=1/2$
 in the nonleptonic kaon decays in the lowest order of
 the ``radiation corrections''.
 Paraphrasing Lord
 Eddington: ``A proton yesterday and electron today do not make an
 atom''~\cite{ed} one can say that ``A kaon
 yesterday and pion today do not make a weak
 $K\to \pi$ transition'' \cite{am,va}.
 The dominance of the instantaneous interaction
 of massive vector boson time components justifies
 the application of the chiral perturbation theory
 leading to meson form factors  \cite{vp1,fpp}.
 Exploiting
 the instantaneous weak interaction (IWI)
 mechanism of enhancement in the
 $K \to \pi$ transition probability
 and QCD symmetry we derive the relation between
 meson form factors  and  the differential
 semileptonic
 decay  rates (\ref{dr}). These relations allow us to estimate
 parameters of $K \to \pi e^+ e^-$ decay  rate in the NA48/2 CERN
 experiment.

  These results show us that the Dirac Hamiltonian formulation \cite{pvn6}
  of Standard Model can reveal new physical effects in the comparison with
  its FP heuristic version \cite{db} used now for the
  describing observational data. Therefore, the problem of
  such the formulation becomes topical in the light of the
  future precision experiments.

\section*{Acknowledgments}

The authors are grateful to
   B.M. Barbashov, D.Yu. Bardin, A.Di Giacomo, S.B. Gerasimov, G.V. Efimov,
    A.V. ~Efremov, V.D. Kekelidze, E.A. Kuraev, A.N. Sissakian
 and M.K. Volkov for fruitful discussions. The work was in part supported by the
 Slovak Grant Agency for Sciences VEGA, Gr.No.2/4099/26.

{}
\end{document}